\documentclass[final,5p,times,twocolumn]{elsarticle}

 \usepackage{graphicx}
 \usepackage{epsfig}

\usepackage{amssymb}
\usepackage{amsmath}

\newcommand{\subfig}[2]{Fig.~\ref{fig:#1}(#2)} %for instance \subfig{LargeDiamonds}{b}
\newcommand\degrees[1]{\ensuremath{#1^\circ}}

\newcommand{\InxGaAs}[2]{\mbox{$\text{In}_{#1}\text{Ga}_{#2}\text{As}$}}
\newcommand{\InAsxP}[2]{\mbox{$\text{In}\text{As}_{#1}\text{P}_{#2}$}}
\newcommand{\Vds}{\mbox{$\text{V}_{\text{ds}}$}}

\newcommand{\WQW}{\mbox{$\text{W}_{\text{QW}}$}}
\newcommand{\tQW}{\mbox{$\text{t}_{\text{QW}}$}}
\newcommand{\Vb}{\mbox{$\text{V}_{\text{b}}$}}
\newcommand{\Vd}{\mbox{$\text{V}_{\text{d}}$}}
\newcommand{\Va}{\mbox{$\text{V}_{\text{a}}$}}
\newcommand{\Vcd}{\mbox{$\text{V}_{\text{c},\text{d}}$}}
\newcommand{\Vc}{\mbox{$\text{V}_{\text{c}}$}}
\newcommand{\Wbase}{\mbox{$\text{W}_{\text{base}}$}}
\newcommand{\Wtop}{\mbox{$\text{W}_{\text{top}}$}}
\newcommand{\Rtwopt}{\mbox{$\text{R}_{\text{2pt}}$}}
\newcommand{\Rc}{\mbox{$\text{R}_{\text{c}}$}}
\newcommand{\Rsheet}{\mbox{$\text{R}_{\text{sh}}$}}

\journal{Physica E}

\begin{document}

\begin{frontmatter}



\title{Electron transport in gated InGaAs and InAsP quantum well wires in selectively-grown InP ridge structures}


\author{G.~Granger}
	\ead{Ghislain.Granger@nrc.ca}	
\author{A.~Kam}
\author{S.A.~Studenikin}
\author{A.S.~Sachrajda}
\author{G.C.~Aers}
\author{R.L.~Williams}
\author{P.J.~Poole}
\address{Institute for Microstructural Sciences, National Research Council, 1200 Montreal Rd., Ottawa, ON, Canada, K1A 0R6}

\begin{abstract}

The purpose of this work is to fabricate ribbon-like InGaAs and InAsP wires embedded in InP ridge structures and investigate their transport properties. The InP ridge structures that contain the wires are selectively grown by chemical beam epitaxy (CBE) on pre-patterned InP substrates. To optimize the growth and micro-fabrication processes for electronic transport, we explore the Ohmic contact resistance, the electron density, and the mobility as a function of the wire width using standard transport and Shubnikov-de Haas measurements. At low temperatures the ridge structures reveal reproducible mesoscopic conductance fluctuations.  We also fabricate ridge structures with submicron gate electrodes that exhibit non-leaky gating and good pinch-off characteristics acceptable for device operation. Using such wrap gate electrodes, we demonstrate that the wires can be split to form quantum dots evidenced by Coulomb blockade oscillations in transport measurements.

\end{abstract}


\end{frontmatter}


\section{Introduction}
\label{intro}

Many solid-state systems have been investigated lately with the goal of forming spin qubits, the basic building block of a quantum computer \cite{Hanson2007}.
Currently, lateral quantum dot devices fabricated from high mobility AlGaAs/GaAs heterostructures using split-gate technology occupy the leading position, as coherent spin manipulations have been demonstrated in this system \cite{koppens:2006,laird:2007,pioro-ladriere:2008}.
However, lateral split-gate technology faces topological obstacles for large scale integration of qubits into a quantum computer, as well as other issues, such as electron spin decoherence due to interaction with nuclear spins.
Therefore, a wide search for different solid-state qubit candidates  continues; for example, it is worth mentioning the triple dot formed in a carbon nanotube \cite{grove-rasmussen:2008}, the double dot formed in silicon \cite{lim:2009}, the etched quantum dot in an InGaAs/InP quantum well (QW) structure \cite{larsson2008}, and the few-electron double quantum dot in InAs/InP nanowire heterostructures~\cite{fuhrer:2007}. Quantum dots made of InGaAs material are of a particular interest because of its large g$^*$-factor ($|$g$^*|$=4.2$\pm$0.2 \cite{dobers:1990}) and its strong spin-orbit coupling (resulting in a spin-splitting energy of about 1~meV at zero magnetic field  \cite{yu:2008}) that is expected to facilitate spin manipulation at high frequencies using local gates.

Selective area growth techniques, which have recently been revived by the interest for quantum information and spintronics applications \cite{akabori:2008}, enable a precise position control of  nanostructures defined by the pattern design, e.g. quantum dots or nanowires, making this technology suitable for achieving scalable circuits. Unlike Molecular Beam Epitaxy (MBE), both Metal-Organic Vapor Phase Epitaxy (MOVPE) and Chemical Beam Epitaxy (CBE) allow selective area epitaxy without the formation of polycrystalline growth on the mask \cite{caenegem:1997}. For CBE there is no lateral transport of source materials across the mask surface \cite{ kayser:1991}, which is not the case for MOVPE where the growth is dependent on the mask layout \cite{tsuchiya:2005}. This can significantly simplify the mask design when growing by CBE. Here we study a new material system for quantum electronics applications: QW wires embedded in InP ridge structures prepared by CBE on pre-patterned substrates. This technology has also been developed for optical single quantum dot devices \cite{reimer:2008}. To make ridge structures conductive,  either InGaAs or InAsP QWs are inserted during the CBE growth. Optimal contact resistance and  electron conductance characteristics obtained from magnetotransport measurements are presented, and gated structures are discussed in terms of current-voltage curves, gating characteristics, and transport measurements in the Coulomb blockade regime.

\begin{figure}[htb]
\setlength{\unitlength}{1cm}
\begin{center}
\begin{picture}(8,4.5)(0,0)
\includegraphics[width=8.0cm, keepaspectratio=true]{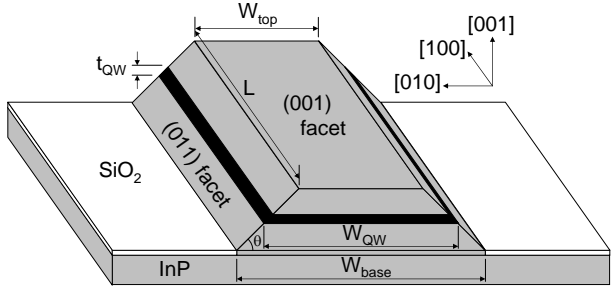}
%\put(0,0){\includegraphics[width=7.0cm, keepaspectratio=true]{chirality.pdf}}
\end{picture}
\end{center}
\caption{Schematic of a ridge. The SiO$_2$ nanotemplate mask is shown in white. Both the substrate and CBE-grown InP are in gray, while the QW is shown in black. Si donors (not shown) are incorporated in the ridge before the QW is grown.}
\label{fig:fig1}
\end{figure}

\begin{figure}[htb]
\setlength{\unitlength}{1cm}
\begin{center}
\begin{picture}(8,6.5)(0,0)
\includegraphics[width=8.0cm, keepaspectratio=true]{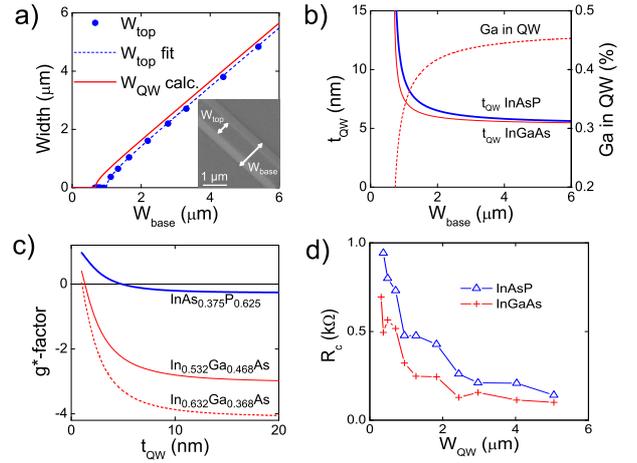}
%\put(0,0){\includegraphics[width=7.0cm, keepaspectratio=true]{chirality.pdf}}
\end{picture}
\end{center}
\caption{(Color online.)
(a) Determination of QW widths from measured ridge base widths. Circles are the ridge top widths \Wtop~as a function of ridge base widths \Wbase, both measured from electron micrographs such as the one shown in the inset. The dashed line is a fit as described in the text, and the solid line is the QW wire width \WQW~calculated with the same fit parameter.
(b) Calculated QW thicknesses \tQW~for both InAsP (thick solid line) and InGaAs (thin solid line) QWs as a fuction of ridge base withs. Shown with a dashed line is the percentage of Ga in InGaAs QWs.
(c) Calculated g*-factor as a function of QW thickness for different QW compositions. The case of InAs$_{0.375}$P$_{0.625}$ is shown as a thick solid line, while those of In$_{0.532}$Ga$_{0.468}$As and In$_{0.632}$Ga$_{0.368}$As are shown as a thin solid line and a dashed line, respectively.
(d) Average ohmic contact resistances extracted from two-point resistance measurements at 4~K, using ridges with L=5 and 10 $\mu$m after illumination with a red LED. Values of \Rc~for InAsP (InGaAs) QWs are shown as triangles (crosses).}
\label{fig:fig2}
\end{figure}

\section{Experimental Details}

The samples studied here are grown by CBE on InP (001) substrates that have a patterned 20~nm thick SiO$_2$ layer on top~\cite{williams2001}. The patterns are made by electron-beam lithography together with a wet etch to create long and narrow openings through the SiO$_2$ to the InP surface. These openings, up to 5~$\mu$m wide, are aligned along the $<$010$>$ directions. During the CBE process, growth occurs on the exposed InP surface and no deposition is observed on the SiO$_2$ surface~\cite{poole:2008}. The growth consists of an InP buffer, Si-doped InP layer, and an InP spacer. This is followed by the growth of the conducting channel, either lattice matched \InxGaAs{0.53}{0.47} or compressively strained \InAsxP{0.375}{0.625}, and then capped with InP. The composition for the InGaAs QW is the nominal value for a layer grown on an unpatterned substrate. When grown on a ridge, the Ga concentration in the layer will drop as the ridge width decreases~\cite{poole:2008}. A schematic of a ridge with inserted QW is shown in Fig.~\ref{fig:fig1}. After the ridges are grown, the SiO$_2$ mask is removed by a wet etch.

The measurements are performed in a $^3$He cryostat equipped with a 5T superconducting magnet in the temperature range between T=0.25 K and 7.5 K. Resistance measurements are made using a standard low-noise resistance bridge with a typical excitation voltage between 30 and 100~$\mu$V, whereas conductance measurements are made with a current pre-amplifier and a lock-in amplifier with a typical 50~$\mu$V excitation voltage. The ribbon-like QW wires  are contacted with NiAuGe pads fabricated by electron-beam lithography and annealed at \degrees{400}C to obtain typical contact resistances below 1 k$\Omega$. In order to form quantum dots in wires, TiAu submicron gate electrodes are patterned on ridges, also using electron-beam lithography.

\section{Results and discussion}

The first step in the characterization of the ridges is to determine the width, \WQW, of the wires within the ridges. This is done as in Ref.~\cite{poole:2008}, using scanning electron microscopy (SEM) to confirm the ridge growth rate. An example of SEM of a ridge is shown in the inset of \subfig{fig2}{a}, with the ridge top width \Wtop~and base width \Wbase~indicated by arrow bars. The data for~\Wtop~for ridges with 0.6$<$\Wbase$<$5.5~$\mu$m are plotted \textit{vs}.~\Wbase~as filled circles in \subfig{fig2}{a}. To describe ridge dimensions including the QW wire, we use the growth model from Ref.~\cite{poole:2008}, which takes into account details of sticking and diffusion characteristics.

The fitted curve is presented as a dashed line in \subfig{fig2}{a}. The values of \WQW~are then calculated. The extracted values of \WQW~are valid not only for the InAsP QWs, but also for the InGaAs structures from the same batch of pre-patterned InP substrates, because the QW wire width is determined by the InP ridge underneath the QW. For very narrow ridges, we find that \WQW~is finite even when \Wtop~is unmeasurably small.   This situation occurs for ridges where the InP cap completes the ridges to a line rather than to a facet with measurable \Wtop.

Using the growth model from Ref.~\cite{poole:2008}, it is also possible to calculate the QW thickness, \tQW, as a function of \Wbase, and the results of this calculation are shown in \subfig{fig2}{b} as a thick (thin) solid line for InAsP (InGaAs) QWs. As the ridge becomes narrower, \tQW~increases.  This is a consequence of the fact that all of the source material that lands on the ridge diffuses to the top surface where it incorporates, with the growth rate on that top surface depending on the ratio of sidewall area to top area. The thickness of the InGaAs QWs increases more slowly than that of the InAsP QWs, as Ga does not diffuse off the sidewalls to the top surface~\cite{poole:2008}. This also means that as the ridges narrow the In composition of the InGaAs ridges increases.

As mentioned earlier, InGaAs and InAsP offer the opportunity for g*-factor engineering.  For example, \subfig{fig2}{c} presents calculation results of electron g*-factor as a function of QW thickness, \tQW~\cite{oestreich1996}.
It is seen that the g*-factor in InAsP/InP is very small and changes sign at a \tQW~of about 5~nm, whereas the g*-factor of InGaAs/InP has much larger negative values and changes sign at much smaller \tQW, in agreement with Ref.~\cite{croke2005}.

The values of \WQW~for a given \Wbase~from \subfig{fig2}{a} can be used to extract the contact resistances of the ridge Ohmic contacts. This is achieved by measuring the two-point resistance \Rtwopt~of ridges with two different lengths (L=5 and 10~$\mu$m) and using the equation \Rtwopt=2\Rc+\Rsheet L/\WQW, where \Rsheet~is the sheet resistance of the two-dimensional electron gas (2DEG) from which the wires are made~\cite{shur1996}. Extrapolating the graph of \Rtwopt~\textit{vs}.~L to L=0 and dividing the answer by 2 gives the value of \Rc~for a given \WQW. The resulting values for the Ohmic contact resistance are all below 1~k$\Omega$ and they decrease as \WQW~increases, as can be seen from the data in \subfig{fig2}{d}. These data are obtained after illumination with a red Light Emitting Diode (LED). In the case of the InAsP QW wires, the illumination improves the contacts by reducing \Rc; for instance, a 14\% (37\%) reduction is observed for \WQW=5.0~$\mu$m (1.0~$\mu$m) (not shown). \Rc~values for InGaAs QW wires in the dark are more difficult to analyze (not shown);  nevertheless, the values of \Rc~after illumination plotted in \subfig{fig2}{d} indicate that the dependence on \WQW~is consistent between InAsP and InGaAs QWs, and the Ohmic contact resistance becomes smaller for wider wires.

\begin{figure}[htb]
\setlength{\unitlength}{1cm}
\begin{center}
\begin{picture}(8,7)(0,0)
\includegraphics[width=8.0cm, keepaspectratio=true]{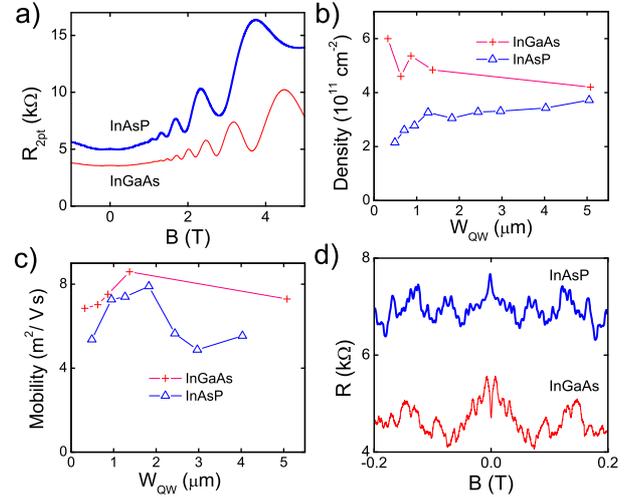}
%\put(0,0){\includegraphics[width=7.0cm, keepaspectratio=true]{chirality.pdf}}
\end{picture}
\end{center}
\caption{(Color online.)
(a) Examples of Shubnikov-de Haas oscillations measured in the two-point resistance \textit{vs}.~perpendicular field B in the dark for an InAsP QW with \WQW$\approx$1.0 $\mu$m at T=6~K (thick solid line) and an InGaAs QW with \WQW$\approx$0.9~$\mu$m at T=4.2 K (thin solid line).
(b) and (c) Density and mobility extracted from Shubnikov-de Haas oscillations such as those in (a) for QWs with 0.3~$\mu$m$<$\WQW$<$5.1~$\mu$m. Results for InGaAs (InAsP) QWs are shown with crosses (triangles).
d) Mesoscopic effects in the ridge resistance versus B at T=0.3 K. The resistance of an InAsP QW with \WQW$\approx$0.6~$\mu$m (thick line) is measured after illumination with a red LED. The data from an InGaAs QW with \WQW$\approx$0.5~$\mu$m in the dark (thin line) are offset for ease of comparison.
}
\label{fig:fig3}
\end{figure}

As discussed above, the QW parameters vary with the ridge width. The transport parameters are also expected to depend on the ridge width. In order to extract the electron density and mobility of the 2DEG inside the ridges, the samples are placed in a perpendicular magnetic field B up to 5~T and the Shubnikov-de~Haas (SdH) oscillations in the two-point resistance are recorded in the dark as the field is swept (see \subfig{fig3}{a}). Figures~\ref{fig:fig3}(b) and (c) show the density and mobility of both InGaAs and InAsP QWs as a function of \WQW. The mobility $\mu$ is calculated from the Drude model as $\mu$=$(\Rsheet n e)^{-1}$, where \Rsheet~is extracted from the resistance at zero magnetic field, $n$ is the electron density, and $e$ is the elementary charge. The mobility values in \subfig{fig3}{c} are comparable to those measured with top-doped planar InGaAs/InP heterostructures in Ref.~\cite{studenikin:2003} and are limited by impurity scattering and not by alloy scattering in the ternary alloy channel~\cite{ramvall:1998}. The mean free path obtained from the expression $\hbar \mu \sqrt{2\pi n}/e$ ranges between 0.8~$\mu$m and 1.0~$\mu$m for the InGaAs QWs and between 0.4~$\mu$m and 0.7~$\mu$m for the InAsP QWs, which makes these ridges suitable for the formation of quantum point contacts with submicron gate electrodes.  Note that the density and mobility can be increased once the ridges are illuminated with a red LED. For instance, the electron density in ridges with InAsP QWs increases by an average of 50\% and their mobility by 20\% (not shown).

The density variation \emph{vs.} wire width shown in \subfig{fig3}{b} have opposite behaviors for InGaAs and InAsP QWs, and this may be due to differences in strain between the two types of QWs. As the base width is reduced in InGaAs samples, there is a competition between a strong increase in density because of the Ga composition change in the QW and a somewhat weaker decrease in density due to the increased QW-doping layer distance and increased QW depth leading to a net increase in density. In the InAsP QW samples there is no change in QW composition with ridge width reduction so the dominant effect should be the QW-doping layer distance, which could lead to a reduction in net density as \WQW~decreases. We have checked that the removal of the QW Ga composition term from the model does indeed flatten the curve from the significantly negative slope in the InGaAs case, but it is still slightly negative. In addition, another factor could be the large strain in the InAsP layer which will relax at the sidewalls. This would lead to a  reduction in band gap from ridge centre to edge. Coupled to a surface pinning mechanism, this could produce a decreasing density as \WQW~decreases. This aspect will need to be investigated further.
%~70meV reduction in band gap

Distinct mesoscopic oscillations are observed in the magnetoresistance of narrow ridges in small magnetic fields as is seen in \subfig{fig3}{d}. Universal Conductance Fluctuations (UCFs) that are symmetric with respect to the magnetic field are seen for both InAsP and InGaAs QWs. In the case of InAsP, a weak localization peak is seen at zero magnetic field, indicating constructive interference between a path of scattering events through the wire and its time-reversal counterpart. In the case of InGaAs, there is a weak anti-localization dip at zero magnetic field originating from the strong spin-orbit coupling in this material \cite{studenikin:2003}.  Detailed analysis of the mesoscopic features is beyond the scope of this publication.

\begin{figure}[htb]
\setlength{\unitlength}{1cm}
\begin{center}
\begin{picture}(8,9.5)(0,0)
\includegraphics[width=8.0cm, keepaspectratio=true]{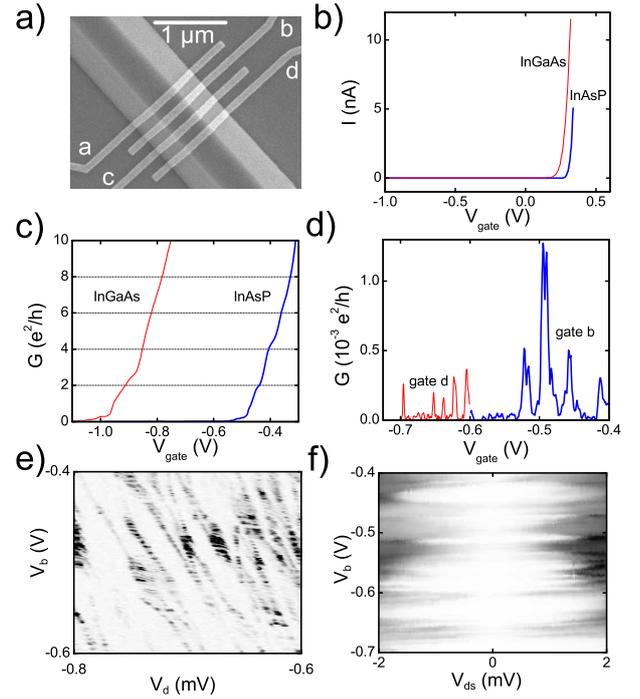}
%\put(0,0){\includegraphics[width=7.0cm, keepaspectratio=true]{chirality.pdf}}
\end{picture}
\end{center}
\caption{(Color online.)
(a) Electron micrograph of a ridge with four submicron Schottky gate electrodes.
(b) Examples of I-V characteristics of the submicron Schottky gate electrodes measured DC in the dark. Turn on is observed in forward bias, where the current flows from the gate electrode to the QW and out through an ohmic contact. Data for an InGaAs (InAsP) QW with \WQW$\approx$0.7~$\mu$m at T=4.2~K (0.6~$\mu$m at T=0.26~K) are shown as a thin (thick) solid line.
(c) Two-point conductance $G$ as a function of submicron gate voltage $V_g$ for both an InGaAs QW with \WQW$\approx$1.0~$\mu$m at T=4~K (thin line) and an InAsP QW with \WQW$\approx$0.6~$\mu$m at T=2~K in the dark with an excitation voltage of 50~$\mu$V (thick line). The resistance at $V_g$=0 has been subtracted from these data.
(d) Dark conductance vs.~voltage on gates d (thin line) and b (thick line) of an InAsP QW with \WQW$\approx$0.6~$\mu$m at T=0.254~K and with 50~$\mu$V excitation voltage.
(e) Stability diagram showing the conductance as a function of both b and d gate voltages at T=0.27~K and with 50~$\mu$V excitation voltage. (\Va,\Vc)=(0.1, -0.6)~V. The conductance goes from 0 (white) to 3.5$\times$10$^{-3}$~e$^2$/h, but all data with G$>$0.001~e$^2$/h are black.
(f) Differential conductance as a function of both the voltage on gate b and ridge drain-source voltage \Vds~at T= 0.27~K with a 100~$\mu$V excitation voltage. (\Va,\Vcd)=(0.25, -0.6)~V. The differential conductance goes from 0 (white) to 2.2 e$^2$/h (black).
}
\label{fig:fig4}
\end{figure}

Conductive ridges can also be gated, in particular, with submicron wrap gates  that are about 100~nm wide (see \subfig{fig4}{a}). Examples of low temperature I-V curves for electrons flowing from the electrode to the QW and out the ohmic contact are shown in \subfig{fig4}{b}. It is seen from this figure that the turn on voltage with an InAsP wire is larger (0.27~V) than for an InGaAs wire (0.16~V). This makes it more difficult to obtain good Schottky gates on InGaAs structures. Large leakage current of InGaAs/InP Schottky barriers prevents the use of standard split gate technology on large 2DEG wafers. So far, only metal-dielectric-semiconductor gated structures have been reported in InGaAs/InP  material~\cite{sun2009}. In our case, the small area of the ridge samples and the wire encapsulation  inside the ridge seem to reduce such leakage problems and render the devices operational without using extra dielectric layers. This is an important advantage of using ridge structures instead of planar wafers.

Each biased finger gate acts as a constriction along the ridge, which can be pinched off if sufficiently negative voltage is applied to the gate electrode. Typical conductance traces for both InGaAs and InAsP QWs are shown in \subfig{fig4}{c}.  Both types of ridge structures can be easily pinched off by the finger gates.  Conductance steps are seen, but they are not quantizatized for the InGaAs QW, most likely, because of a voltage dependent series resistance. The InAsP QW reveals steps that are approximately quantized in increments of 2e$^2$/h. In order to obtain clear quantized characteristics, we have modified the gate geometry, and these newly designed samples are currently under test. Quantized conductance steps were observed in GaAs/AlGaAs wires grown on patterned substrates \cite{pfaendler:2008}, in etched GaAs/AlGaAs wires \cite{kasai:2002}, and in etched GaInAs/InP wires \cite{martin:2008}. The wires in these three examples were short (the wire length was shorter than the mean free path), which explains the observation of quantized steps.

A quantum dot can be formed if several gate electrodes are biased simultaneously. Indeed, Coulomb blockade oscillations are seen when sweeping either gate d or b, as shown in \subfig{fig4}{d}. The stability diagram that corresponds to this situation for the conductance in the \Vb-\Vd~plane is shown in \subfig{fig4}{e}. The Coulomb charging peaks are seen as diagonal lines. However, these lines are chopped by quasi-horizontal white lines. These may come from incidental quantum dots forming in the sidewall of the ridge, right under the finger gates. Figure~\ref{fig:fig4}(f) shows the standard diamond plot measurements for the differential conductance in the \Vb-\Vds~plane at \Vd=-0.6 V, where \Vds~is the drain-source voltage. The size of the diamonds in the \Vds~direction gives an addition energy between 1 and 2 meV.

\section{Conclusion}

In conclusion, we have fabricated ribbon-like QW wires in InP ridge structures using state-of-the-art selective growth epitaxy and explored their transport properties.  We have demonstrated the suitability of such structures for the fabrication of gated quantum dot devices. In particular, the Ohmic contacts have resistances below 1 k$\Omega$, and the ridges have good transport characteristics.  Measured electron density and mobility are in the same range as obtained for planar wafers. Gated ridges can be pinched off without noticeable gate leakage and tuned to the Coulomb-blockade regime. In the future, we plan to improve our gate designs  and eventually reach the few-electron regime for high frequency spin and quantum states manipulation experiments.

The authors acknowledge L. Gaudreau and P. Zawadzki for discussions and experimental help. G.G.~acknowledges financial support from the CNRS-CNRC collaboration and A.S.S.~acknowledges support from CIFAR.

\end{document}